\documentclass[useAMS,usenatbib]{mn2e}
\usepackage{epsfig}
\usepackage{graphicx}

\begin{document}

\title[NS spin-kick alignment and DNS coalescence rates]
{Neutron star spin-kick velocity correlation effect on 
binary neutron star coalescence rates and spin-orbit misalignment of the components}

\author[K.A. Postnov, A.G. Kuranov]
{K.A. Postnov, A.G. Kuranov\\
{\sl Sternberg Astronomical Institute, Universitetski pr. 13,
Moscow, 119991, Russia}
\thanks{E-mail: pk@sai.msu.ru(KAP); alex@xray.sai.msu.ru(AGK)}}

\date{Accepted ......  Received ......; in original form ......
      }

\maketitle

\begin{abstract}
We study the effect of the neutron star spin -- kick velocity
alignment observed in young radio pulsars on the coalescence rate of binary neutron stars. Two scenarios of the neutron star formation are considered: when the kick is always
present and when it is small or absent if a neutron star is formed in
a binary system due to electron-capture degenerate core collapse. 
The effect is shown to be especially strong for large kick amplitudes and tight alignments, reducing
the expected galactic rate of binary neutron star coalescences
compared to calculations with randomly directed kicks. The spin-kick correlation also
leads to a much narrower NS spin-orbit misalignment.
\end{abstract}


\section{Introduction}
There is an increasing interest of a broad astrophysical community
to coalescing binary compact stars as primary sources of
gravitational waves for the ground-based gravitational wave observatories. Double neutron stars (DNS), observed
as binary pulsars, remain to be the most reliable objects for gravitational wave searches. On-going LIGO science runs
\citep{Abbot07} have already set first experimental upper limits on their galactic rates of a few per year. Astrophysical estimates of the
DNS coalescence rate, which are based on the binary pulsar statistics or can be obtained from population synthesis simulations,
are model-dependent and vary within more than an order of magnitude around the value $10^{-5}$ per year (see recent reviews \citealt{PYu,Kalogera&07} and references therein).

The kick velocity imparted to a newborn neutron star is an important phenomenological parameter of the core collapse supernovae and represents one of the major uncertainties in the theory of binary star evolution.
The origin of the kicks remains unclear and a number of physical models have been suggested (see, for example, \citealt{Lai}
and references therein). For post-supernova evolution of a binary, both
the amplitude of the kick and its space direction are important. The distribution of the kick amplitudes is usually obtained from the analysis of radio pulsar proper motions \citep{Hobbs}. The
direction of kicks (for example, with respect to the spin axis of the neutron star) is more difficult to infer from observations.
Recently, several observational clues
appeared indicating possible NS spin-kick alignment. A noticeable spin-kick alignment
has been inferred from polarization measurements of radio emission of pulsars \citep{Johnston_ea, Rankin07, Johnston_ea07},
as well as from X-ray observations
of pulsar wind nebulae around young pulsars \citep{Helfand,Kargaltsev}.
Implications of these findings to the formation of double
pulsars were discussed by \cite{Wang_ea06}.
The possibility and conditions for such an alignment in the model
of the kick origin by multiple random kicks
during NS formation (proposed by \citep{Spruit}) were studied by \cite{Wang_ea07}. 
The implication of NS kick-spin correlation to the plausible birth-kick scenarios was also discussed by \cite{Ng&Romani}.

Here we explore the effect of NS spin -- kick correlation
on the formation and galactic coalescence rate of double neutron stars (DNS) which
are primary targets for modern gravitational wave detectors. We show that the tighter alignment,
the smaller is the DNS merging rate with respect to models with random kick orientation. The effect
is especially important for large kick amplitudes ($\sim 400$ km/s). We calculate the spin-orbit misalignment of the components of DNS which can be important for GW data analysis.
We also considered a scenario in which no (or insignificant) kick accompanies the
formation of a neutron star in binary systems from the main-sequence progenitors in a restricted
mass range (8-11 $M_\odot$ or so) due to electron-capture collapse of O-Ne-Ng degenerate
stellar core proposed by \cite{Podsiadlowski_ea04} and further elaborated by \cite{vdH04,vdH07}. This hypothesis
is phenomenologically based on the existence of 
long-period Be X-ray binaries with low eccentricities
\citep{Pfahl_ea02}.
It is consistent with the evolutionary analysis of double neutron star formation
\citep{vdH07} and has been used in some population synthesis studies of DNS, see for example \cite{Dewi_ea05, Dewi_ea06}.


\section{Effect on the binary neutron star coalescence rates}

The effect of NS kick velocity on merging rates of compact binaries by means of population synthesis simulations was
studied earlier (e.g. \cite{Lipunov_ea97, PZYu, BelczKalogera01, Belcz_ea02}) and the inclusion of the
kick into modeling of binary star evolution is a prerequisite in all
population synthesis simulations (see \citealt{PYu,Kalogera&07} for discussion and further references).
Clearly, the tight NS spin -- kick alignment may have important implications to
the formation and evolution of binary compact stars, as was shown previously by \cite{Kalogera}.

Consider the standard evolutionary scenario leading to the formation of a
binary NS from a massive binary system \citep{B&vdH91},
which is also discussed in the review \cite{PYu},
focusing
on the effect of the NS kick velocity\footnote{Here we do not consider other possible evolutionary scenario for DNS formation, e.g. \protect{\cite{Brown_95}}, which allows for hypothetical hypercritical accretion onto NS in the common envelope. These scenarios are studied by means of the population synthesis modeling by other authors \protect{\citep{Dewi_ea05, Dewi_ea06}}}.
We shall assume that the kick velocity vector
is confined within a cone which is coaxial with the progenitor's rotation axis and
characterized by angle $\theta<\pi/2$. We shall consider only central kicks thus ignoring
theoretically feasible off-center kicks simultaneously affecting the NS spin~\citep{Spruit,Postnov&Prokhorov,Wang_ea07}.
The value of the kick velocity is assumed to obey
the Maxellian distribution $f(v)\sim v^2 \exp(-(v/v_0)^2)$, as suggested by pulsar
proper motion measurements \citep{Hobbs}. In our analysis we varied
the velocity $v_0$ from 0 to 400 km/s.

\begin{figure*}
\begin{minipage}{\textwidth}
\psfig{figure=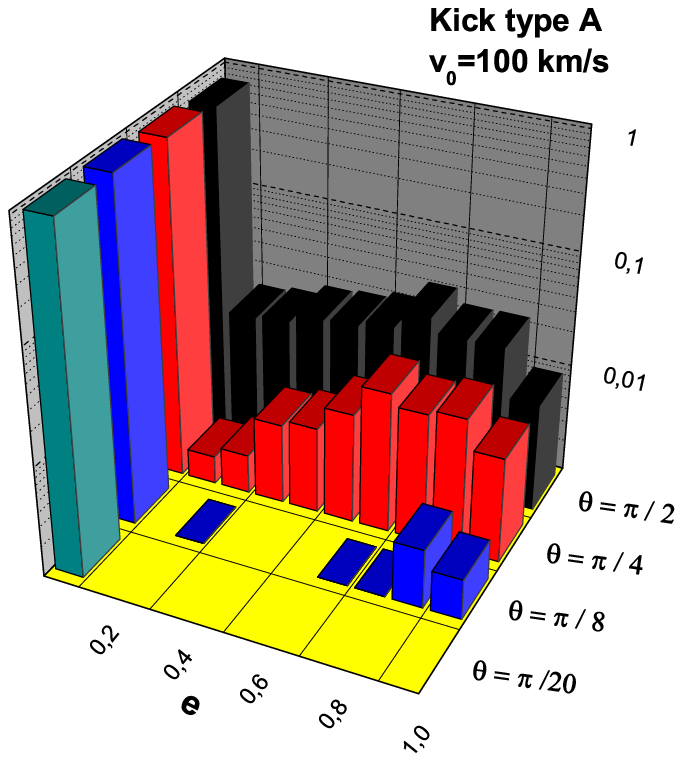,width=0.48\textwidth}
\psfig{figure=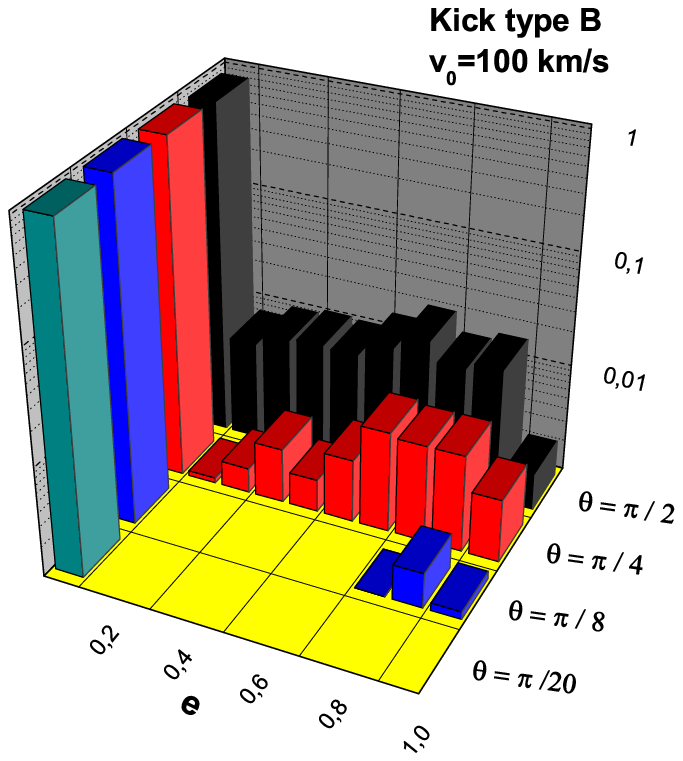,width=0.48\textwidth}
\hfill
\caption{The distribution of orbital eccentricities
before the second collapse in binaries producing DNS
for two kick models. Kick type A: all NS in binaries receive a kick; kick type B: the NS kick is zero in those binaries where NS is produced
from main-sequence progenitors with masses $8-11 M_\odot$.}
\label{f:ecc}
\end{minipage}
\end{figure*}

The rotational axes of both components are assumed to
be aligned with the orbital angular momentum before the primary collapses to form the first NS.
The SN explosion is treated in a standard way as instantaneous
loss of mass of the exploding star.
The effect of the kick on the post-explosion binary orbital parameters is treated using the energy-momentum conservation in the two point-mass body problem
(see the description in e.g. \citealt{Hills83,Kalogera,Grishchuk}). The first SN explosion most likely occurs when the binary orbit is circular (unless the initial binary is very wide so that tidal circularization is ineffective), while the second explosion can
happen before the orbit has been tidally circularized.
Possible mass transfer phases before the second collapse
(such as the common envelope stage and
stable mass transfer onto neutron star) are assumed to effectively circularize the orbit. In the absence of mass transfer the tidal evolution of the orbit eccentricity is treated according to \cite{Zahn_77}.
In our modeling, by the time of the second collapse the fraction of eccentric binaries which later form DNS
 attains $\sim 10\%$ depending on the kick velocity
value and direction, as illustrated in Fig. \ref{f:ecc}.
It is higher for isotropic kicks and increases with their
absolute values. To treat the explosion in an eccentric
binary we choose the
position of the star in the orbit randomly distributed according to Kepler's 2d law.

\begin{figure*}
\psfig{figure=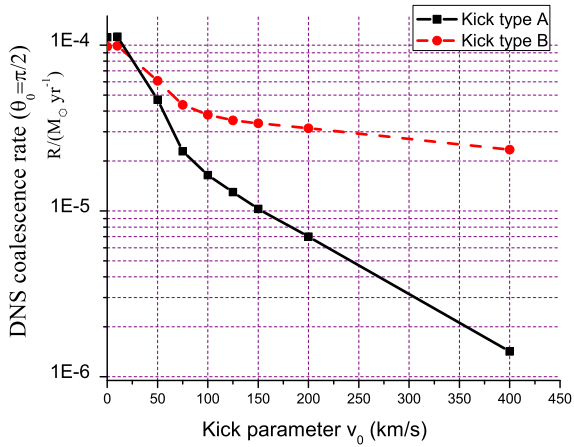,width=0.48\textwidth}
\caption{Galactic coalescence rate of DNS vs. the kick parameter $v_0$ assuming random central kicks. Almost an exponential decay with $v_0$ is seen for $v_0>100$ km/s for kick type A (a), while the decrease in the rate is smaller for kick type B}.
\label{f:coalrate}
\end{figure*}

\begin{figure*}
\begin{minipage}{\textwidth}
\psfig{figure=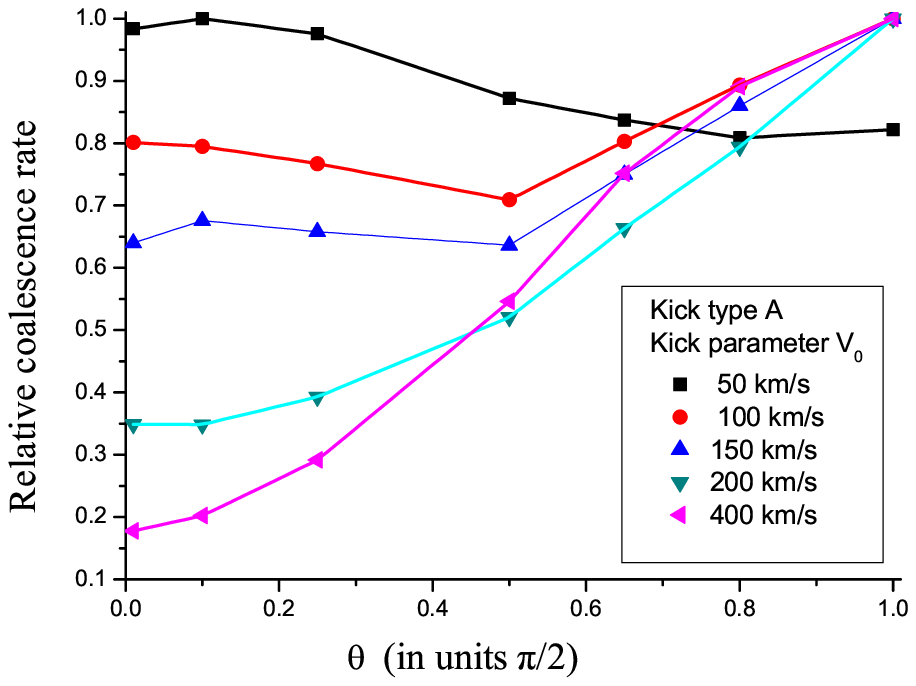,width=0.48\textwidth}
\psfig{figure=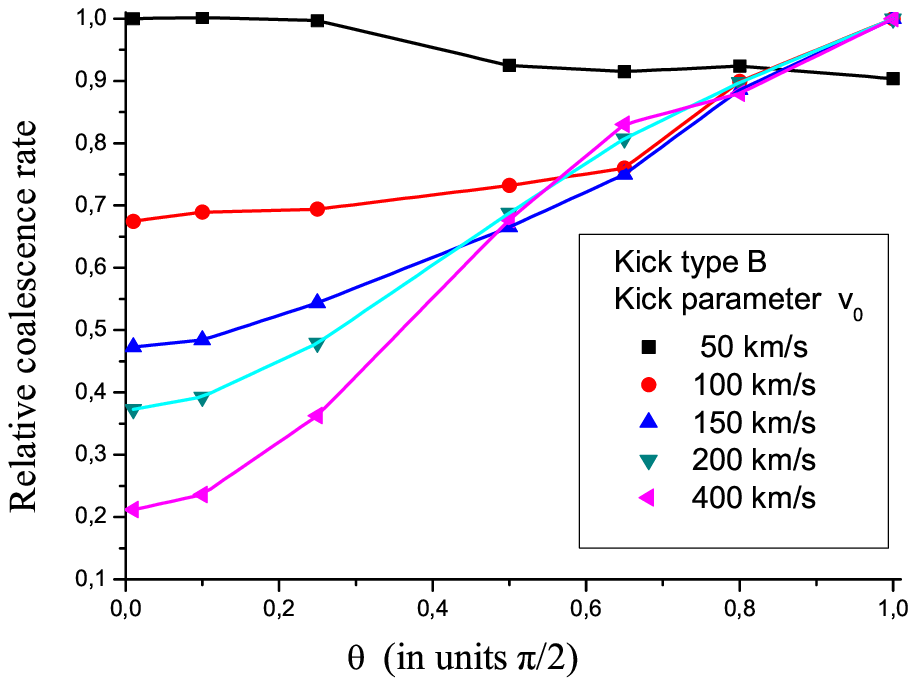,width=0.48\textwidth}
\hfill
\caption{Relative change of DNS merging rate for the NS spin-kick correlation as a function of the kick restriction angle $\theta$ for two kick models.}
\label{f:relrate}
\end{minipage}
\end{figure*}

We use the population synthesis method to calculate the expected coalescence rate
of DNS (see \citep{Lipunov_ea97,PYu} and references therein). The standard assumptions about
binary evolution have been made: Salpeter's mass function for the primary's mass,
$dN/dM_1\sim M_1^{-2.35}$,
a flat initial mass ratio ($q=M_2/M_1<1$) distribution $dN/dq=const$, the initial semi-major axes distribution 
in Oepik's form $dN/d\log a=const$. The common envelope phase is treated in the standard way based on the energy conservation \citep{PYu} with the efficiency $\alpha_{CE}=0.5$\footnote{This important parameter of the evolution of close binaries is loosely constrained, see e.g. the detailed discussion in \protect\cite{PYu}; however, varying it from 0.1 to 1 does not change qualitatively the shape of  distributions studied in the present paper.}.
The calculations were normalized to the galactic star formation rate
$3M_\odot$ per year, with a binary fraction of 50\%. 
The maximum mass of a main sequence star which forms a NS in the collapse is set to 30 $M_\odot$, the maximum mass of a NS is assumed to be
2 $M_\odot$. No hypercritical accretion onto NS, as assumed to be possible in the scenario by Brown (1995), is allowed. 
We also have carefully taken into account rotational evolution
of magnetized compact stars, as described in detail in \cite{Lipunov,LPP96},
assuming no neutron star magnetic field decay. 

The galactic DNS merging rate
is shown in Fig. \ref{f:coalrate} as a function of the kick parameter $v_0$
and assuming random central kicks.
The calculations were performed for two assumptions about kicks --
(a) when the formation of NS is always accompanied by a kick
(we refer to this scenario as kick type A) or (b) when
the kick is non-zero during the formation of NS only in binaries
starting out from $11 M_\odot$, while there is no kick velocity at all when a NS is formed in a binary system due to the electron-capture core collapse of the main-sequence progenitor with mass in the range 8-11$M_\odot$ 
(kick type B). In the case (a)
an almost exponential decrease in
the DNS rate coalsecence rate
with $v_0$ for $v_0>100$ km/s is seen. In the case (b) the decrease with $v_0$ is less pronounced, mainly because kickless collapses in binaries
are more abundant by the assumed Salpeter mass distribution.

Fig. \ref{f:relrate} shows the relative change
in the DNS merging rate with allowance for the NS spin-kick alignment with different values of the
kick confinement angle $\theta$ for two kick models.
It is seen that tight alignment (small $\theta$)
generally reduces the DNS merging rate,
with the effect being especially strong for large kick velocity amplitudes. Such a decrease relative to
calculations with random kicks is clear because the NS spin -- kick correlation
excludes kicks in the binary orbital plane which, if directed opposite to the orbital velocity,
can additionally bind the post-explosion binary system.
\begin{figure*}
  \begin{minipage}{\textwidth}
\psfig{figure=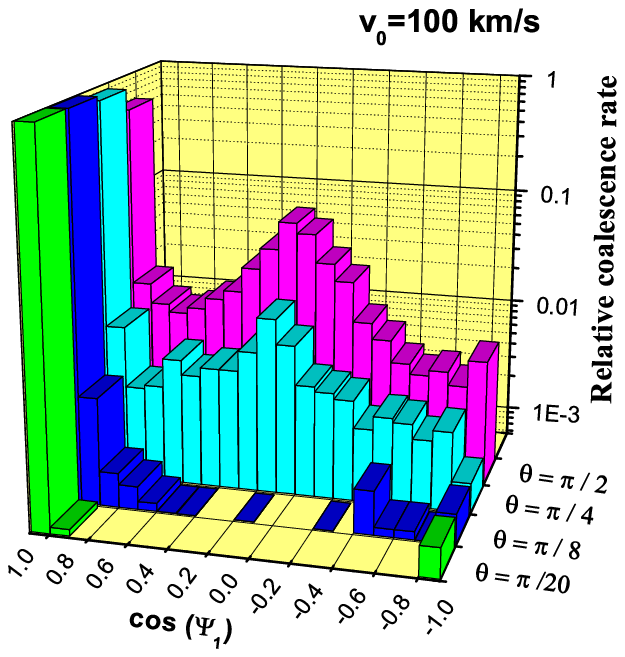,width=0.5\textwidth}
\psfig{figure=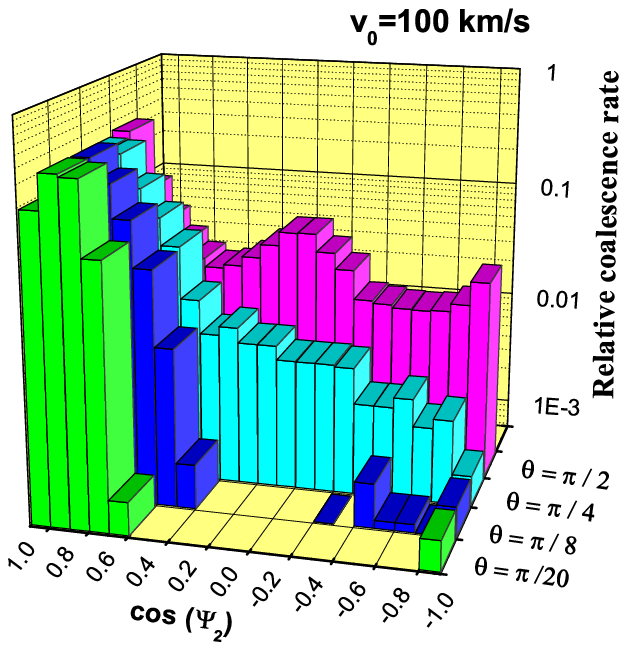,width=0.5\textwidth}
\hfill
\psfig{figure=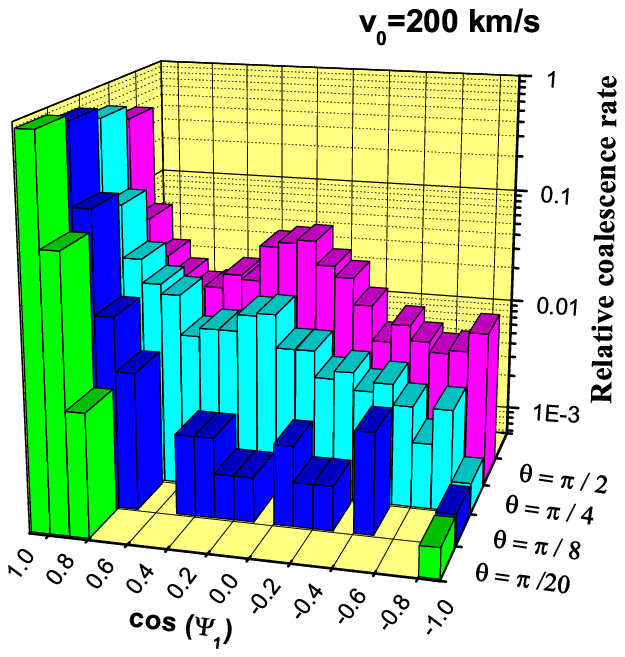,width=0.5\textwidth}
\psfig{figure=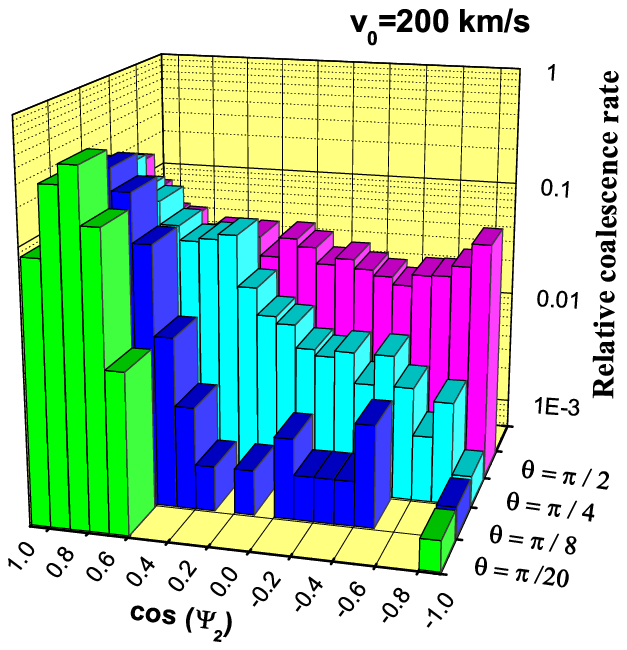,width=0.5\textwidth}
\hfill
\psfig{figure=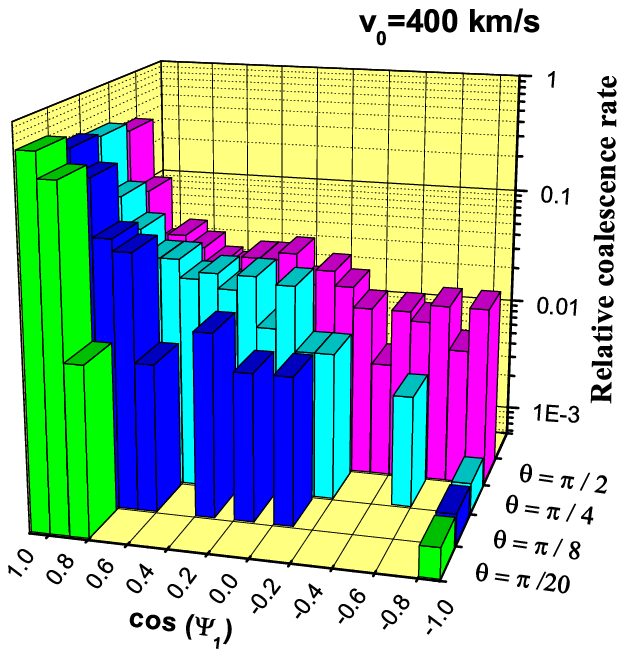,width=0.5\textwidth}
\psfig{figure=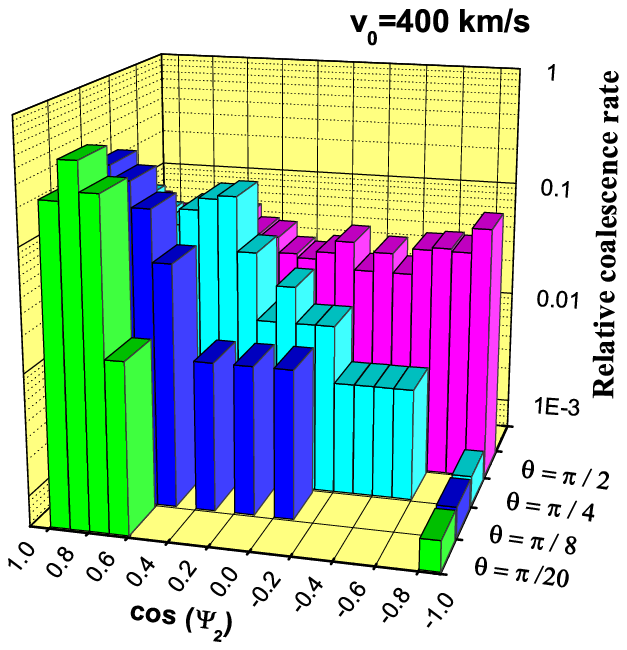,width=0.5\textwidth}
\hfill
\caption{NS spin-orbit misalignment ($\cos\Psi$) in coalescing DNS for kick type A with $v_0=100$ km/s (upper row),  $v_0=200$ km/s (middle row)
 and $v_0=400$ km/s (bottom row) and different NS spin-kick alignment angles $\theta$.
Left panels: the NS1 and secondary component's spins aligned with the orbital angular momentum
prior to the SN2 explosion (the case of close binaries). Right panels: the NS1 and secondary component's spins aligned with the original binary's orbital angular momentum prior to the SN2 explosion.} 
\label{f:mis_a}
\end{minipage}
\end{figure*}

\begin{figure*}
  \begin{minipage}{\textwidth}
\psfig{figure=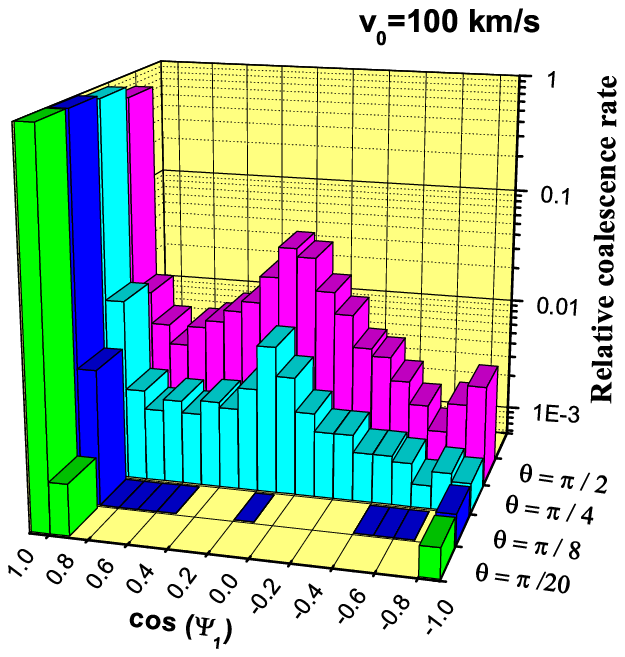,width=0.5\textwidth}
\psfig{figure=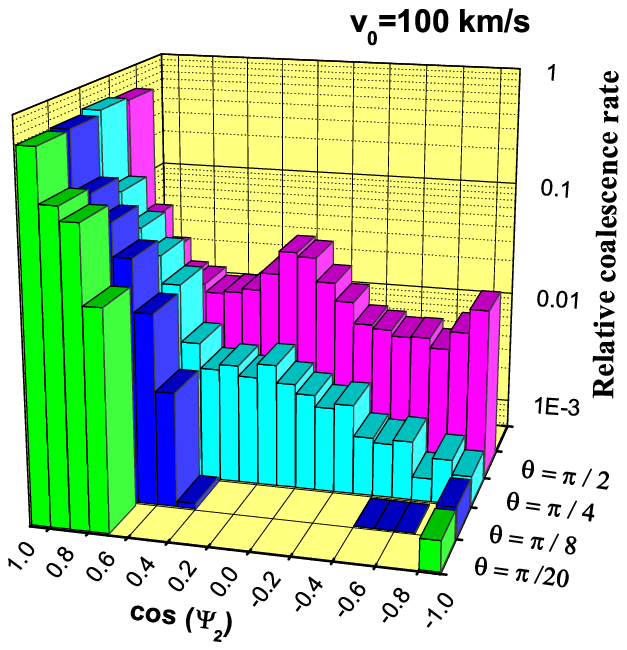,width=0.5\textwidth}
\hfill
\psfig{figure=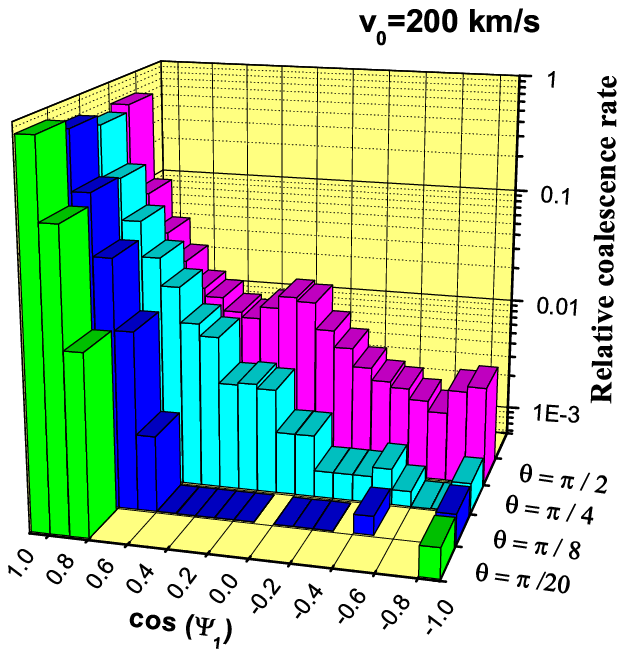,width=0.5\textwidth}
\psfig{figure=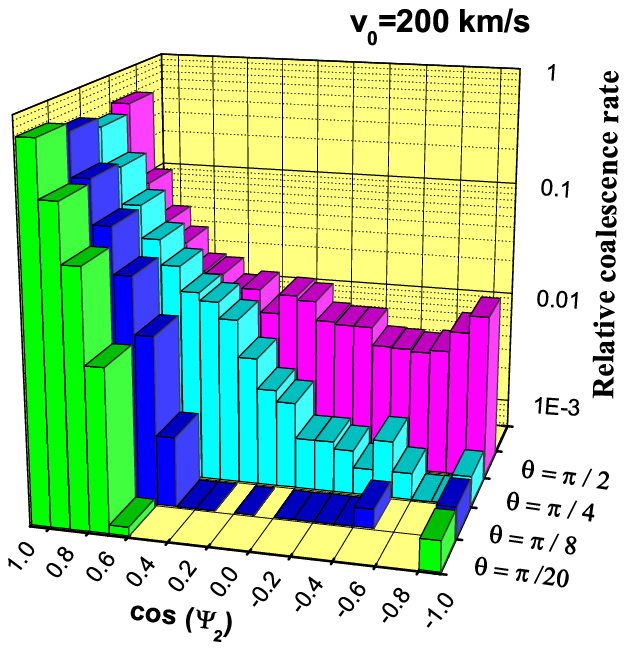,width=0.5\textwidth}
\hfill
\psfig{figure=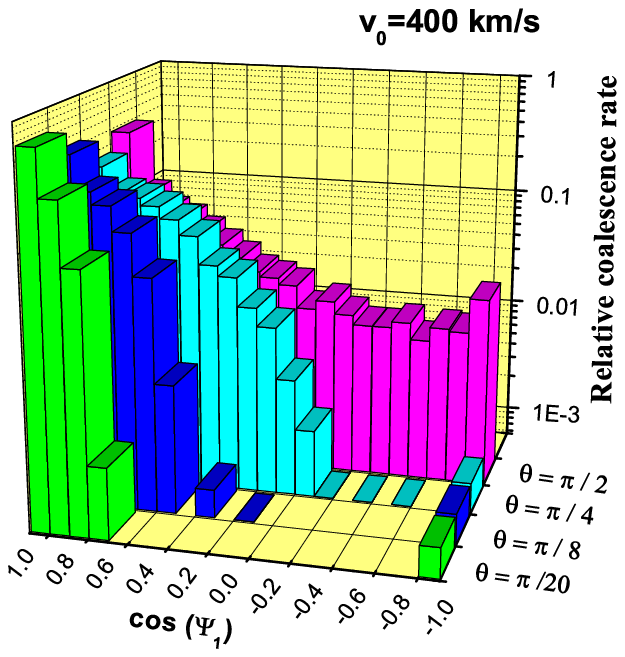,width=0.5\textwidth}
\psfig{figure=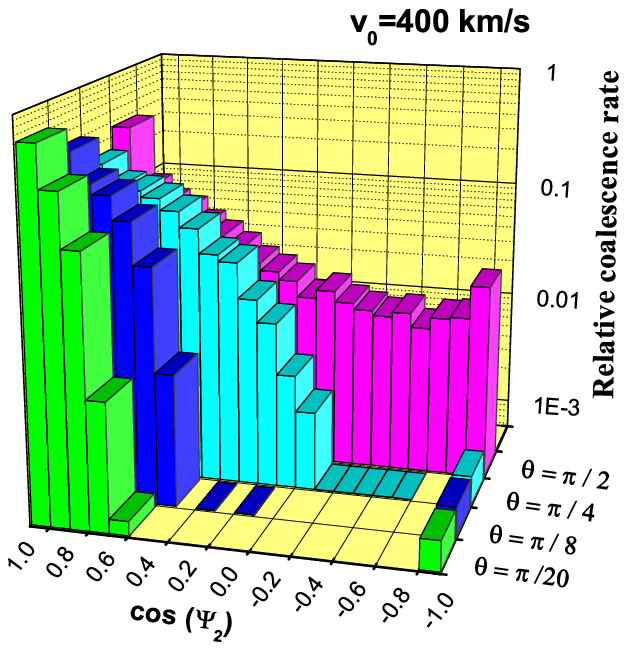,width=0.5\textwidth}
\hfill
\caption{The same as in Fig. \protect{\ref{f:mis_a}} for kick type B.}
\label{f:mis_b}
\end{minipage}
\end{figure*}

\section{Neutron star spin -- orbit misalignment}

There is another observational
consequence of the kick in DNS systems: the NS spin -- orbit misalignment, which can be tested by geodetic precession measurements in binary pulsars \citep{Bailes}. Such a misalignment
is potentially very interesting for GW studies \citep{Apostolatos}.
After the first supernova explosion in a binary system (SN1), the additional kick imparted to the newborn neutron star (NS1)
results, in general, in a misalignment between
the new orbital angular momentum and the NS1 (as well as the secondary component's)
spin vector characterized by some angle. For the instantaneous 
explosion of one of the components treated as point-like masses on a Keplerian
orbit, this angle can be calculated analytically, see for example \cite{Kalogera}. After the second supernova explosion in the system (SN2), there are several possibilities for spin-orbit misalignment of the compact components.

1) In close binaries, tidal interactions tend to rapidly align the angular momentum
vector of the normal star with the orbital angular momentum. To spin-up the
NS rotation to observed ms periods (in binary ms pulsars), a modest amount of matter ($\sim 0.1 M_\odot$) should
be accreted by NS. This amount is sufficient to align the NS rotation with the orbital angular momentum.
So if NS1 accreted matter before the second SN explosion,
both NS1 and the secondary component's spins should be most likely aligned with the orbital
angular momentum (see the discussion in \cite{Wang_ea06}). Note that the NS1 spin tends to
align with the orbital angular momentum even if NS1 does not accrete matter
but spins-down by the propeller mechanism before the second SN explosion, since in that case very strong currents must flow through its polar cap and the alignment torque can be as strong as during accretion.
So in close binaries the NS1 remains orbit-misaligned prior to the second SN explosion only in rare occasions where the secondary collapses shortly after the first SN in the binary.
If both NS1 and the secondary component were aligned with orbital angular momentum prior to the second SN explosion, both neutron stars NS1 and NS2 will be equally misaligned with the orbital angular momentum ($\Psi_1$)
after SN2 with a kick.

2) In sufficiently wide binaries, when
tidal interactions between the components are inefficient, the orientation of the NS1 spin
and the secondary's spin vector may remain unchanged until SN2 explosion, after which
the orbital angular momentum vector changes again due to the NS2 kick. So in this case we would
expect two coaxial NS with spins misaligned by angle $\Psi_2$ with orbital angular momentum.
However, such binaries, unless highly eccentric, may be too wide
to coalesce over the Hubble time.

We conclude from these considerations that the components of a DNS can have coaxial spins misaligned with the orbit by angles $\Psi_1$ or $\Psi_2$ depending on the strength of tidal interaction 
(weak or strong, respectively) acting between two
SN explosions in the binary system. It is of course possible that spins of the components remain misaligned by some angle depending
on the degree of the spin-orbit interaction of the secondary prior to the collapse. For example, 
NS1 spin may conserve its original direction in space, while the secondary component before the collapse 
may have become aligned with the orbital angular momentum, so in the resulting DNS 
the spin-orbit misalignment angle of the older NS will be $\Psi_2$ while that of the younger NS will be $\Psi_1$.
In this sense, angles $\Psi_1$ and $\Psi_2$ should be considered as limiting cases.

In our population synthesis simulations
we take into account the discussed spin alignment effects.
In Fig. \ref{f:mis_a} and
\ref{f:mis_b} we show calculated distributions between spins of the components and
the orbital angular momentum in coalescing DNS systems
assuming tidal spin-orbit alignment before the second collapse (angle $\Psi_1$) and the conservation of the original components spin direction before SN2 (angle $\Psi_2$).

It is seen that the misalignment angles
can be very different (and even with negative cosines) for random or loosely constrained
($\theta \sim \pi/2$) kicks, while a tight spin-kick
alignment ($\theta\ll \pi/2$) results in much narrow distributions (see also \cite{Kalogera}).
The mean spin-orbit misalignment angles $\Psi$ for different values of the kick velocity parameter $v_0$ and kick models A and B
are presented in Table \ref{t:a} and \ref{t:b}, respectively. Fig. \ref{f:mis_a}, \ref{f:mis_b} and Tables \ref{t:a}, \ref{t:b} also show that the difference
between $\Psi_1$ and $\Psi_2$ (and, hence, between spin-orbit misalignment of NS1 and NS2 in the
final pre-merging DNS) can be appreciable for small kicks, but tends to be
less significant for high kick velocities.

\begin{table}
\caption{Mean NS spin-orbit misalignment $\Psi$ (in units $\pi/2$) for kick type A.}
\bigskip
\begin{center}
\begin{tabular}{llccccc}
\hline
$v_0$,&\multicolumn{6}{c}{Kick confinement angle $\theta$ (in units $\pi/2$)}\\
km/s&           &\bf{ 0.01} & \bf{0.1} & \bf{0.25} & \bf{0.5}& \bf{1.0} \\
\hline
\bf{ 50}&$\Psi_1$ & 0.061 & 0.059 & 0.057 & 0.071 & 0.208\\
        &$\Psi_2$ & 0.307 & 0.312 & 0.321 & 0.305 & 0.335\\
\bf{100} &$\Psi_1$& 0.109 & 0.108 & 0.113 & 0.187 & 0.438\\
 	 &$\Psi_2$& 0.378 & 0.380 & 0.394 & 0.425 & 0.629\\
\bf{200} &$\Psi_1$& 0.190 & 0.192 & 0.221 & 0.311 & 0.503\\
         &$\Psi_2$& 0.420 & 0.421 & 0.456 & 0.568 & 0.816\\
\bf{400} &$\Psi_1$& 0.251 & 0.257 & 0.316 & 0.399 & 0.547\\
         &$\Psi_2$& 0.454 & 0.462 & 0.521 & 0.671 & 0.959\\
\hline
\label{t:a}
\end{tabular}
\end{center}
\end{table}

\begin{table}
\caption{Mean NS spin-orbit misalignment $\Psi$ (in units $\pi/2$) for kick type B.}
\bigskip
\begin{center}
\begin{tabular}{llccccc}
\hline
$v_0$,&\multicolumn{6}{c}{Kick confinement angle $\theta$ (in units $\pi/2$)}\\
km/s&           &\bf{ 0.01} & \bf{0.1} & \bf{0.25} & \bf{0.5}& \bf{1.0} \\
\hline
\bf{ 50}&$\Psi_1$ &  0.039   &   0.039  &   0.039   &   0.054   &   0.170   \\
        &$\Psi_2$ &  0.252   &    0.254  &   0.260   &  0.251   &    0.270\\
\bf{100} &$\Psi_1$&  0.088   &   0.090  &   0.092   &   0.139   &   0.284 \\
 	 &$\Psi_2$&  0.246   &    0.250  &   0.254    &  0.265   &    0.391\\
\bf{200} &$\Psi_1$  &  0.167   &   0.177  &   0.216   &   0.282   &   0.296   \\
         &$\Psi_2$   &  0.214   &    0.222  &   0.260    &  0.329   &    0.368\\
\bf{400} &$\Psi_1$  &  0.162   &   0.191  &   0.296   &   0.446   &   0.510   \\
         &$\Psi_2$    &  0.176   &    0.202  &   0.308 & 0.458   &    0.530\\
\hline
\label{t:b}
\end{tabular}
\end{center}
\end{table}

The NS spin-orbit misalignment can be probed by studying radio pulsars in binary systems. An extensive analysis of observational data was done by \cite{Wang_ea06}. However, high uncertainties in the inferred spin-orbit misalignment angles have not allow firm conclusions to be made as yet.

\section{Conclusions}

We have shown that the spin-velocity correlation observed in radio pulsars,
suggesting the NS spin-kick velocity alignment, may have important implications
to GW studies. First, the tight alignment reduces the galactic rate
of double neutron star coalescences (especially for large kicks 300-400 km/s)
relative to models with random kicks. Second, the spin-kick correlation results in a
specific distribution of NS spin -- orbit misalignments. In turn, analysis of the NS spin-orbit misalignments inferred from GW signals during DNS mergings can be potentially used to put independent bounds on the still elusive nature of NS kicks.

\section*{Acknowledgments}
The authors acknowledge M.E. Prokhorov and S.B. Popov for discussions, and the anonymous referee for useful
comments.
This work was partially supported by the RFBR grant 07-02-00961.


\end{document}